\documentclass[showpacs,aps,prb,amsmath,amssymb,twocolumn,superscriptaddress]{revtex4}
\usepackage{graphicx}
\usepackage{dcolumn}
\usepackage{bm}
\begin{document}

\renewcommand{\thefigure}{\arabic{figure}}
\title{Effect of disorder on the ground-state properties of graphene}
\author{R. Asgari}
\affiliation{School of Physics, Institute for Studies in
Theoretical Physics and Mathematics, 19395-5531 Tehran, Iran}
\author{M. M. Vazifeh}
\affiliation{Department of Physics, Sharif university of Technology, Tehran 11155-9161, Iran}
\author{M. R. Ramezanali}
\affiliation{Department of Physics, Sharif university of Technology, Tehran 11155-9161, Iran}
\author{E. Davoudi}
\affiliation{Department of Physics, Islamic Azad university , Tehran
14168-94351, Iran}
\author{B. Tanatar}
\affiliation{Department of Physics, Bilkent University, Bilkent,
06800 Ankara, Turkey}

\begin{abstract}
We calculate the ground-state energy of Dirac electrons in graphene
in the presence of disorder. We take randomly distributed charged
impurities at a fixed distance from the graphene sheet and surface
fluctuations (ripples) as the main scattering mechanisms. Mode-coupling
approach to scattering rate and random-phase approximation for
ground-state energy incorporating the many-body interactions and the disorder effects yields good agreement with experimental
inverse compressibility.
\end{abstract}
\pacs{73.63.-b  72.10.-d  71.55.-i  71.10.-w}
\maketitle

\section{Introduction}

Two-dimensional crystals of Carbon atoms (graphene) are recently
discovered~\cite{novoselov}. Graphene, a single, one-atom thick
sheet of carbon atoms arranged in a honeycomb lattice. High
quality graphene single crystals some thousands of $\mu m^2$ in
size are sufficient for most fundamental physics
studies.~\cite{geim} There are significant efforts to grow
graphene epitaxially~\cite{berger} by thermal decomposition of
SiC, or by vapor deposition of hydrocarbons on catalytic metallic
surfaces which could later be etched away leaving graphene on an
insulating substrate.

This stable crystal has attracted considerable attention because
of its unusual effective many-body properties~\cite{yafis},
quasi-particle properties and its Landau Fermi liquid
picture~\cite{polini} and the effect of electron-electron interactions to
plasmon behavior and angle resolved photoemission spectroscopy (ARPES)
~\cite{polini2} that follow from chiral band states and because of
potential applications. The low energy quasi-particle excitations
in graphene are linearly dispersing, described by Dirac
cones at the edges of the first Brillouin zone. It is very hard
for alien atoms to replace the carbon atoms in the graphene
structure because of the robustness and specificity of the $\sigma$
bonding. Due to that, electron mean-free path $l$, in graphene can
be very large. One of the important issues in graphene is its
quantum transport properties having the universal minimum
conductivity at the Dirac point. Initially, it was believed that
this universality is a native property~\cite{novoselov2} but
recent experimental\cite{tan,cho} and
theoretical\cite{adam,ando2,pereira,katsnelson2,castro-neto} reports
indicate that the transport properties are very sensitive to
impurities and defects and minimum conductivity is not universal.

Conventional two-dimensional electron gas (2DEG) has been a
fertile source of surprising new physics for more than four decades.
Although the exploration of graphene is still at an early
stage, it is already clear~\cite{novoselov2} that the strong field
properties of Dirac electrons in graphene are different from
and as rich as those of a semiconductor heterojunction 2DEG. The
Fermi liquid phenomenology of Dirac electrons in
graphene~\cite{polini, polini2} and conventional
2DEG~\cite{asgari2} have the same structure, since both systems are
isotropic and have a single circular Fermi surface. The strength of
interaction effects in a conventional 2DEG increases with decreasing
carrier density. At low densities, the quasiparticle weight $Z$ is
small, the velocity is suppressed~\cite{asgari2}, the charge
compressibility changes sign from positive to negative\cite{asgari},
and the spin-susceptibility is strongly enhanced~\cite{asgari2}.
These effects emerge from an interplay between exchange interactions
and quantum fluctuations of charge and spin in the 2DEG.

In the Dirac electrons in graphene, it was
shown~\cite{yafis,polini,polini2} that interaction effects also
become noticeable with decreasing density, although more slowly,
that the quasiparticle weight $Z$ tends to larger values, that the
velocity is enhanced rather than suppressed, and that the
influence of interactions on the compressibility and the
spin-susceptibility changes sign. These qualitative differences
are due to exchange interactions between electrons near the Fermi
surface and electrons in the negative energy sea and to interband
contributions to Dirac electrons from charge and spin
fluctuations.

Compressibility measurements of conventional 2DEG
have been carried out~\cite{eisenstein} and it is found
qualitatively that Coulomb interactions affect the
compressibility at sufficiently low electron density or strong
coupling constant region. Recently, the local compressibility of
graphene has been measured ~\cite{martin} using a scannable single
electron transistor and it is argued that the measured
compressibility is well described by the kinetic energy
contribution and it is suggested that exchange and correlation
effects have canceling contributions. From the theoretical point
of view, the compressibility was first calculated by Peres {\it et
al.}~\cite{peres} considering the exchange contribution to the
noninteracting doped or undoped graphene flake. A related quantity
$\partial\mu/\partial n$ (where $\mu$ is the chemical potential
and $n$ is the electron density) is recently considered
by Hwang {\it et al}.\cite{hwang_dmu} within the same approximation.
Going beyond the exchange contribution, the correlation effects
were taken into account by Barlas
{\it et al}~\cite{yafis} based on an evaluation of graphene's
exchange and random phase approximation (RPA) correlation
energies. Moreover, Sheehy and Schmalian~\cite{sheehy} by
exploiting the proximity to relativistic electron quantum critical
point, derived explicit expressions for the temperature and
density dependence of the compressibility properties of graphene.
All these theoretical efforts have been carried out for clean systems.
Since disorder is unavoidable in any material, there has been
great interest in trying to understand how disorder affects the
physics of electrons in material science specially here in
graphene and its transport properties.

Our aim in this work is to study the ground-state properties in
the presence of electron-impurity and electron-electron
interactions. For this purpose, we use the self-consistent theory
of G{\"o}tze \cite{gotze} to calculate the scattering
rate, ground-state energy and the compressibility of the system at
the level of RPA including disorder effects. Our calculation is
in the same spirit of our earlier work on conventional 2DEG.\cite{asgari}
We note that recent work of Adam {\it et al}.\cite{adam} also
use a self-consistent approach where the impurity
scattering by the charge carriers is treated self-consistently in
the RPA and the static conductivity is calculated in the Boltzmann
kinetic theory. Thus, the main difference between the present
work and that of Adam {\it et al}.\cite{adam} is that we are interested
in a thermodynamic quantity (compressibility) whereas the latter
is aimed at calculating a transport property (conductivity).
We also remark that direct solution of Dirac equation for Dirac-like
electrons incorporating the charge impurities has been discussed
by Novikov~\cite{novikov} and the validity of the Born
approximation is seriously questioned. Similar work has been carried
out by Pereira {\it et al}.\cite{pereira} in which they studied the
problem of a Coulomb charge and calculated the local density
of state and local charge by solving the Dirac equation.
They found new characteristics of bound states and strong
renormalization of the van Hove singularities in the lattice
description that are beyond the Dirac equation.

In this work, we consider the charged impurity and the
surface-roughness potentials which are established
experimentally~\cite{meyer,ishigami} to be important. It has been
demonstrated that a short-range scattering potential is irrelevant
for electronic properties of graphene~\cite{katsnelson,adam}. We
have used the same method ~\cite{asgari, asgari3} to investigate
some properties of the conventional 2DEG. In this paper, we point
out the differences between the graphene and conventional 2DEG due
to disorder effects. The scattering rate behavior within our
self-consistent theory shows that impurity scattering cannot
localize the carriers in graphene. The effect of disorder on spin
susceptibility is similar to that on compressibility and
accordingly we will not show any result for spin susceptibility.

The rest of this paper is organized as follows. In Sec.\,II, we
introduce the models for self-consistent calculation of impurities
effect. We then outline the calculation of compressibility.
Section III contains our numerical calculations of ground state
properties and comparison of models with recent experimental
measurements. We conclude in Sec.\,IV with a brief summary.

\section{Theoretical Model}

We consider a system of 2D Dirac-like electrons interacting via the
Coulomb potential $e^2/\epsilon r$ and its Fourier transform
$v_q=2\pi e^2/(\epsilon q)$
where $\epsilon$ is the background dielectric constant.
The Dirac electron gas Hamiltonian on a graphene sheet is given by
\begin{equation}\label{ham}
{\hat {\cal H}} = v\sum_{\bf k, \alpha} {{\hat \psi}_{{\bf k}, \alpha}}^\dag \left[\tau^3\otimes {\bf \sigma \cdot
k}\right] {\hat \psi}_{{\bf k}, \alpha}  + \frac{1}{2A}\sum_{{\bf
q}\neq 0}v_q ({\hat n}_{\bf q} {\hat n}_{-{\bf q}}-{\hat N})
\end{equation}
where $v=3 t a/2$ is the Fermi velocity, $t$ is the tight-binding
hopping integral, $a$ is the spacing of the honeycomb lattice, $A$
is the sample area and ${\hat N}$ is the total number operator.
Here $\tau^3$ is a Pauli matrix that acts on $K$ and $K'$
two-degenerate valleys at which $\pi$ and $\pi^*$ bands touch and
$\sigma^1$ and $\sigma^2$ are Pauli matrices that act on
graphene's pseudospin degrees of freedom.

A central quantity in the theoretical formulation of the many-body
effects in Dirac fermions is the dynamical polarizability tensor
$\chi^{(0)}({\bf q},i\Omega,\mu\neq 0)$ where $\mu$ is chemical
potential. This is defined through the one-body noninteracting
Green's functions.\cite{gonzalez_1994} The density-density
response function $\chi^{(0)}({\bf q},\Omega,\mu)$ of the doped
two-dimensional Dirac electron model was first consider by
Shung~\cite{shung} as a step toward a theory of collective
excitations in graphite. The Dirac electron $\chi^{(0)}({\bf
q},\Omega,\mu)$ expression has been considered recently by
us~\cite{yafis} and others.~\cite{others} Implementing the
Green's function $G^{(0)}({\bf k},\omega,\mu)$ in the calculation
a closed form expression for $\chi^{(0)}({\bf q},i\Omega,\mu\neq
0)$ is found.\cite{yafis} To describe the properties of Dirac
electrons we define a dimensionless coupling constant
$\alpha_{gr}=g{e^2/\upsilon \epsilon \hbar}$ where $g=g_vg_s=4$ is
the valley and spin degeneracy.

The effect of disorder is to dampen the charge-density fluctuations
and results to modify the dynamical polarizability tensor.
Within the relaxation time approximation the modified
$\chi^{(0)}({\bf q},i\Omega,\mu,\Gamma)$ is given by~\cite{mermin}
\begin{equation}
\chi^{(0)}({\bf q},i\Omega,\mu, \Gamma)= \frac{\chi^{(0)}({\bf
q},i\Omega+i\Gamma,\mu)}{1- \frac{\Gamma}{\Omega+\Gamma}\left[1-
\frac{\chi^{(0)}({\bf q},i\Omega+i\Gamma,\mu)}{\chi^{(0)}({\bf q})}
\right]}~,
\end{equation}
in which the strength of damping is represented by $\Gamma$. To
include the many-body effects, we consider the density-density
correlation function within the RPA,
\begin{equation}
\chi_{\rho\rho}({\bf q},i\Omega,\mu,\Gamma)=\frac{\chi^{(0)} ({\bf
q},i\Omega,\mu, \Gamma)} {1-v_q\chi^{(0)}({\bf q},i\Omega,\mu,
\Gamma)}~.
\end{equation}

As the short-range disorder is shown~\cite{adam} to have
negligible effect in the transport properties of graphene, we
consider long-ranged charged impurity scattering and surface
roughness as the main sources of disorder. The latter mechanism
also known as ripples comes either from thermal fluctuations or
interaction with the substrate.\cite{nima} The disorder averaged
surface roughness (ripples) potential (SRP) is modeled as
\begin{equation}
\langle
|U_{surf}(q)|^2\rangle = \pi\Delta^2h^2 (2\pi e^2 n/\epsilon)^2
e^{-q^2\Delta^2/4} \, ,
\end{equation}
where $h$ and $\Delta$  are parameters
describing fluctuations in the height and width, respectively. We
can use the experimental results of Meyer {\it al}.\cite{meyer}
who estimate $\Delta\sim 10$\,nm and $h\sim 0.5$\,nm. It is
important to point out that there are other models to take into
account the surface-roughness potential. The effect of bending of the
graphene sheet has been studied by Kim and Castro Neto~\cite{kim}.
This model has two main effects, firstly the decrease of the
distance between carbon atoms and secondly a rotation of the $p_z$
orbitals. Due to bending the electrons are subject to a potential
which depends on the structure of the graphene sheet.
Another possible model is described by
Katsnelson and Geim ~\cite{katsnelson} considering the change of
in-plane displacements and out-of-plane displacements due to the
local curvature of a graphene sheet. Consequently, the change
of the atomic displacements results to change in nearest-neighbour
hopping parameters which is equivalent to the appearance of a random
gauge field described by a vector potential. These different models
need to be implemented in our scheme and to be checked numerically
to assess their validity in comparison to the available measurements.

The charged disorder potential (CDP) is taken to be
\begin{equation}
\langle |U_{imp}(q)|^2\rangle=n_i v_q^2 e^{-2qd}\, ,
\end{equation}
in which $n_i$ is the
density of impurities and $d$ is the setback distance from the
graphene sheet.

We use the mode-coupling approximation introduced by
G{\"o}tze\cite{gotze} to express the total scattering rate
in terms of the screened disorder potentials
\begin{displaymath}
i\Gamma=-\frac{v_F k_F }{2 \hbar n A}\sum_{\bf q}
\left[\frac{<\mid U_{imp}(q)\mid^2>}{\varepsilon^2(\bf
q)}\right.
\end{displaymath}
\begin{equation}
\left. +\frac{\langle|U_{surf}(q)|^2\rangle}{\varepsilon^2(\bf
q)}\right]\frac{\varphi_0({\bf q},i\Gamma)}{1+i\Gamma\varphi_0({\bf q},i\Gamma)/\chi^{0}({\bf q})}~,
\end{equation}
where $\varepsilon({\bf q})=1-v_q \chi^{(0)}({\bf
q})$ is the static screening function and the relaxation function for electrons
scattering from disorder is given as
$\varphi_0({\bf q},i\Gamma)=
\left[\chi^{(0)}({\bf q},i\Gamma,\mu)-\chi^{(0)}({\bf q})\right]/i\Gamma$ .

Since the scattering rate $\Gamma$ depends on the relaxation function
$\varphi_0({\bf q},i\Gamma)$, which itself is determined by the disorder
included response function, the above equation needs to be
solved self-consistently to yield eventually the
scattering rate as a function of the coupling constant. Note that at the present
level of approximation (i.e. RPA) the static dielectric function
$\varepsilon(q)$ does not depend on $\Gamma$. In the conventional 2DEG
correlation effects beyond the RPA (through the local-field factor)
render $\varepsilon(q)$ also $\Gamma$ dependent.\cite{asgari}

The ground state
energy is calculated using the coupling constant integration
technique, which has the contributions $E^{tot}=E_{kin}+E_{\rm x}
+E_{\rm c}$.

\begin{figure}[ht]
\begin{center}
\tabcolsep=0 cm
\includegraphics[width=0.75\linewidth]{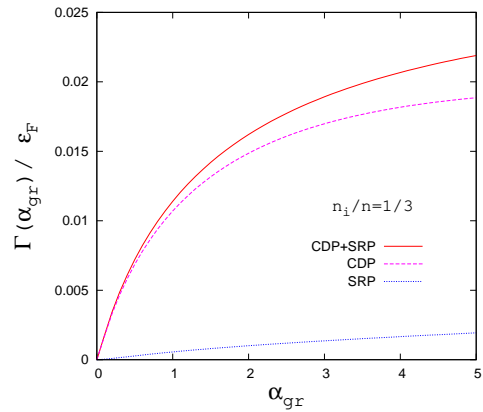}
\caption{(Color online) The scattering rate $\Gamma$ as a function
of the coupling constant $\alpha_{gr}$ for both the
charge-disorder potential (CDP) and surface roughness potential
(SRP) contributions. }
\end{center}
\end{figure}

The first-order ``exchange" contribution per particle is given by
\begin{displaymath}\label{ex}
\varepsilon_{\rm x}=\frac{E_{\rm x}}{N}=\frac{1}{2}\int \frac{d^2
{\bf q}}{(2\pi)^2}~v_q
\end{displaymath}
\begin{equation}
\left[-\frac{1}{\pi n} \int_0^{+\infty}d
\Omega ~\chi^{(0)}({\bf q},i\Omega,\mu,\Gamma)-1\right]\,.
\end{equation}
To evaluate the correlation energy in the RPA, we follow a standard
strategy for uniform continuum models ~\cite{Giuliani_and_Vignale}
\begin{displaymath}\label{corr}
\varepsilon^{\rm RPA}_{\rm c}=\frac{E_{\rm c}}{N}= \frac{1}{2\pi
n}\int \frac{d^2 {\bf q}}{(2\pi)^2}
\int_0^{+\infty}d\Omega\left\{v_q\chi^{(0)}({\bf
q},i\Omega,\mu,\Gamma)\right.
\end{displaymath}
\begin{equation}
\left. +\ln{\left[1-v_q\chi^{(0)}({\bf
q},i\Omega,\mu,\Gamma)\right]}\right\}\,.
\end{equation}

Since $\chi^{(0)}({\bf q},\Omega,\mu,\Gamma)$ is linearly
proportional to ${\bf q}$ at large ${\bf q}$ and decrease only like
$\omega^{-1}$ at large $\omega$, accordingly the exchange and
correlation energy built by Eqs.~(7) and (8) is
divergent~\cite{yafis}. In order to improve convergence, it is
convenient at this point to add and subtract $\chi^{(0)}({\bf
q},i\Omega,\mu= 0,2\Gamma)$ inside the frequency integral and
regularize~\cite{note} the
exchange and correlation energy. Therefore, these ultraviolet
divergences can be cured calculating
\begin{equation}\label{exchange_regularized}
\delta \varepsilon_{\rm x}=-\frac{1}{2\pi n}\int \frac{d^2 {\bf
q}}{(2\pi)^2}~v_q \int_0^{+\infty}d \Omega ~\delta \chi^{(0)}({\bf
q},i\Omega,\mu,\Gamma)
\end{equation}
and
\begin{displaymath}\label{eq:regularization}
\delta \varepsilon^{\rm RPA}_{\rm c}=\frac{1}{2\pi n} \int \frac{d^2
{\bf q}}{(2\pi)^2} \int_0^{+\infty}d\Omega\left\{v_q \delta
\chi^{(0)}({\bf q},i\Omega,\mu,\Gamma)\right.
\end{displaymath}
\begin{equation}
\left. +
\ln{\left[\frac{1-v_q\chi^{(0)}({\bf q},i\Omega,\mu,\Gamma)}{1-
v_q\chi^{(0)}({\bf q},i\Omega,\mu=0,2\Gamma)}\right]}\right\}
\end{equation}
where $\delta \chi^{(0)}$ is the difference between the doped
($\mu \ne 0$) and undoped ($\mu=0$) polarizability functions. With
this regularization the $q$ integrals have logarithmic ultraviolet
divergences~\cite{yafis}. we can introduce an ultraviolet cutoff
for the wave vector integrals $k_c=\Lambda k_F$ which is the order
of the inverse lattice spacing and $\Lambda$ is dimensionless
quantity. Fermi momentum is related to density as given by
$k_F=(4\pi n/g)^{1/2}$. Once the
ground state is obtained the compressibility $\kappa$ can easily
be calculated from
\begin{equation}
\kappa^{-1}=n^2\frac{\partial^2 (n \delta\varepsilon_{\rm tot})}
{\partial n^2}\,,
\end{equation}
where the total ground-state energy is given by
$\delta \varepsilon_{\rm tot}=\delta \varepsilon_{\rm kin}+
\delta \varepsilon_{\rm x}+\delta \varepsilon^{\rm RPA}_{\rm c}$.
Here the zeroth-order kinetic contribution to the ground-state energy
is $\delta \varepsilon_{\rm kin}
=\frac{2}{3}\varepsilon_{\rm F}$. we consider the dimensionless
ratio $\kappa/\kappa_0$ where $\kappa_0=2/(n\varepsilon_{\rm F})$
is the compressibility of the noninteracting system.

\begin{figure}[ht]
\begin{center}
\includegraphics[width=0.7\linewidth]{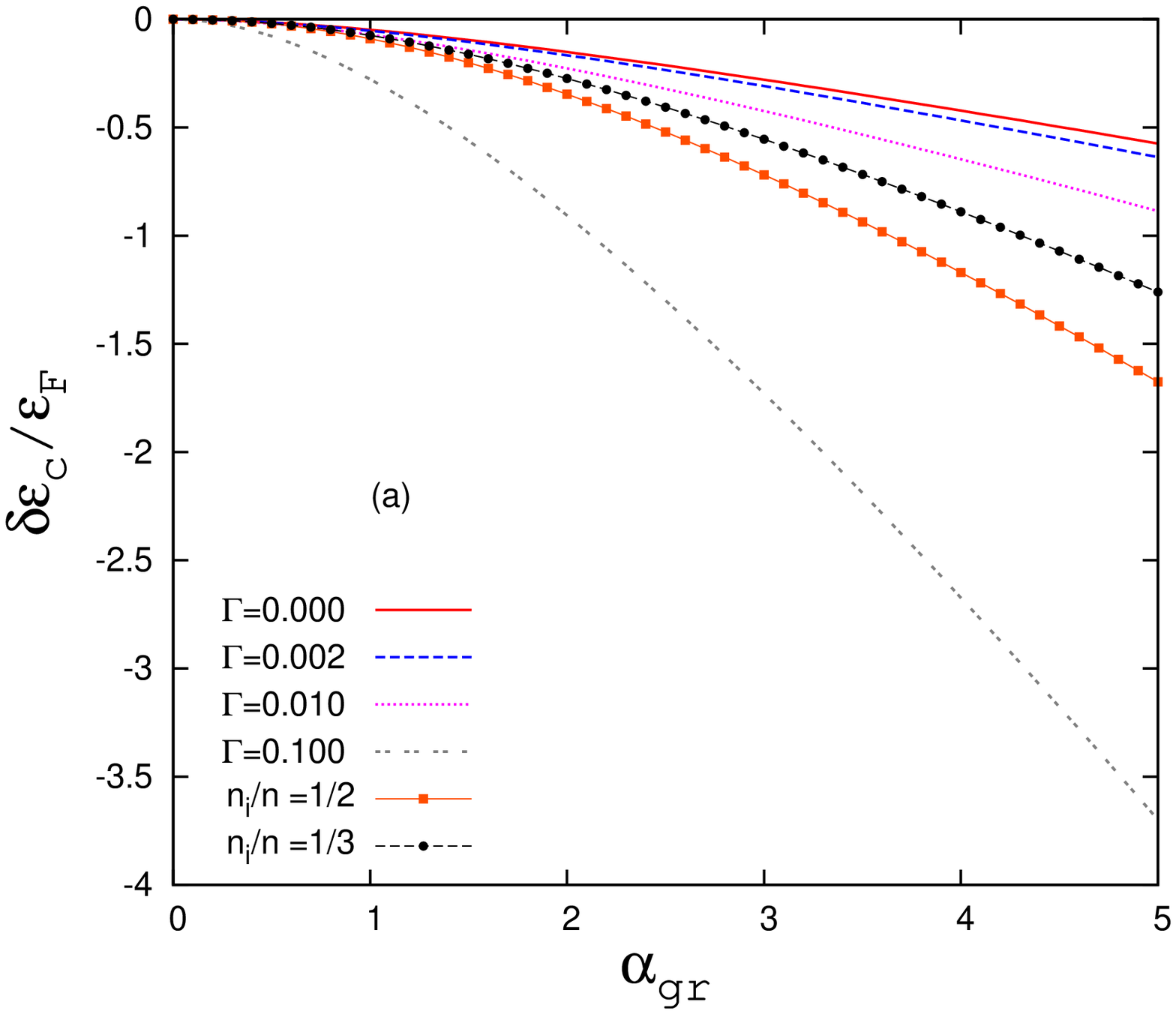}
\includegraphics[width=0.7\linewidth]{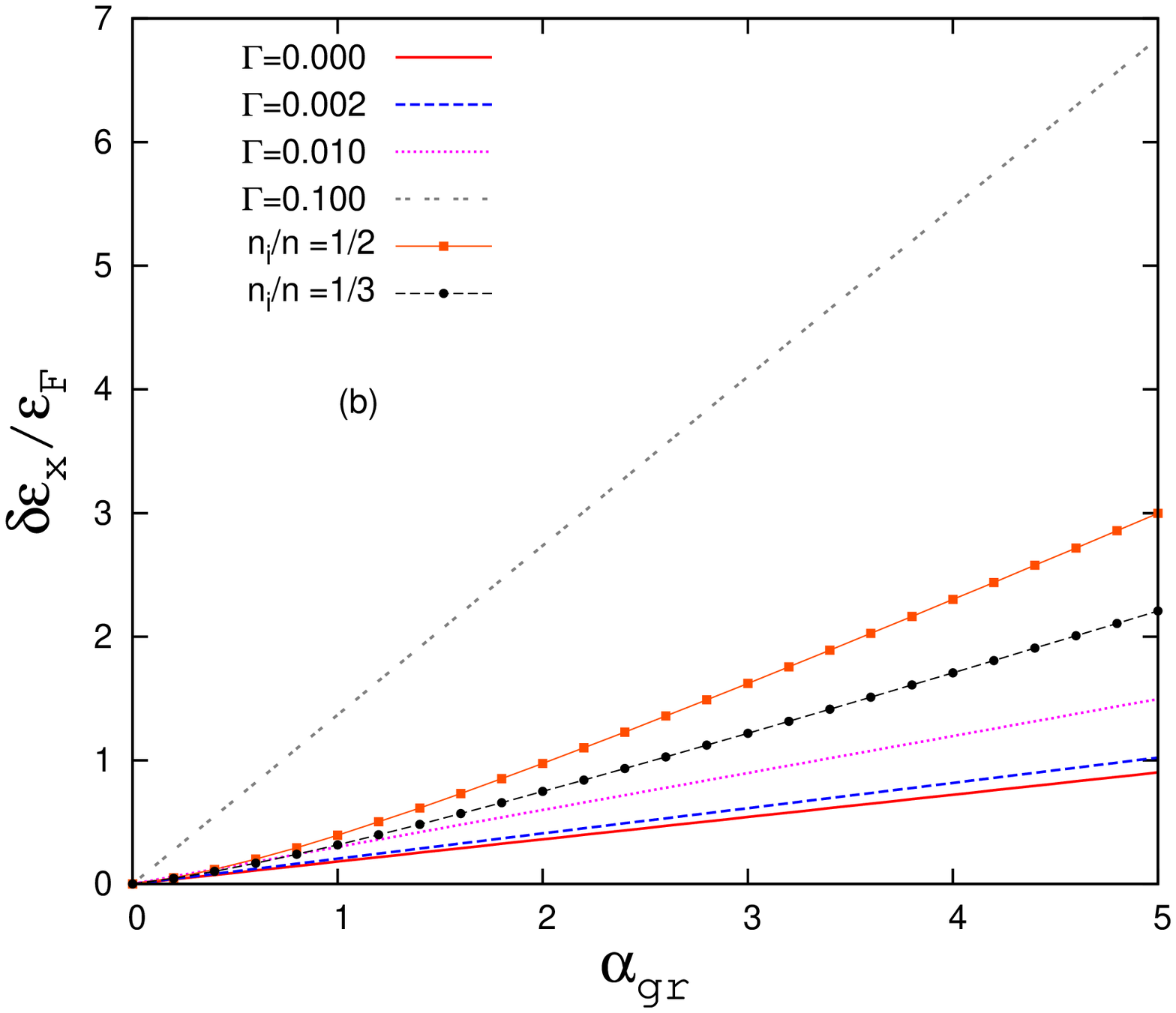}
\caption{(Color online) (a): The correlation energy
$\delta \varepsilon_c$ as a function of the coupling constant
$\alpha_{gr}$ for cut-off value $\Lambda=k_c/k_F=50$.
(b): The exchange energy $\delta \varepsilon_x$ as a function
of the coupling constant $\alpha_{gr}$ for cut-off value
$\Lambda=50$. Results of fixed $\Gamma$ values are compared to
those calculated within the mode-coupling approximation.}
\end{center}
\end{figure}

\section{Numerical results}

In this section we present our calculations for ground state
properties of graphene in present of impurities that we model them
as mentioned above. The inverse compressibility $1/(n^2 \kappa)$
is calculated by using the theoretical models described above and
compare them with the recent experimental measurements.
In all numerical calculations we consider $d=0.5$\,nm. Electron density
is taken to be $1\times 10^{12}$\,cm$^{-2}$ for Figs.\,1-3.

Increasing disorder (increasing $n_i$ or
decreasing $d$ for charge-disorder potential or increasing $h$ for
surface roughness potential) decrease the
$\chi^{(0)}(q,\Omega,\mu,\Gamma)$ as the scattering rate $\Gamma$
gets bigger. Thus, decreasing $\chi^{(0)}(q,\Omega,\mu,\Gamma)$
(or increasing correlation effects) results in a stronger disorder
potential. Despite $\Gamma$ increases with increasing
$\alpha_{gr}$, apparently it turns to a saturation limit and does
not diverge. This behavior is quite different than what is seen in
conventional 2DEG ~\cite{asgari} when the many-body effects
influence the scattering rate through the local-field factor. In
the conventional 2DEG system, at a critical level of disorder this
nonlinear feedback causes $\Gamma$ to increase rapidly and diverge,
which is taken as an indication for the localization of carriers. However,
in graphene, our calculations show that the $\Gamma$ does not diverge
therefore
impurities cannot localize carriers and we have a weakly localized
system in the presence of impurities compatible with experimental
observations~\cite{morozov}. We can understand the saturated behavior
of $\Gamma$ qualitatively as follows.
In the context of conventional 2DEG, Mott argument says that
the mean-free path $l$ in a metal could not be shorter than the wavelength
$\lambda$. Since $l$ is proportional to the inverse of $\Gamma$,
for large values of $\Gamma$ obtained in 2DEG,
the electron mean-free path decreases and becomes less than or equal to
$\lambda$. At this point we should have a metal-insulator phase transition.
In the context of graphene, on the other hand, Mott's argument is
similar to that light does not notice any roughness (one source of
scattering) on a scale shorter than its wavelength.
Consequentely there is a lower limit for the electron's mean-free path
in graphene and it turns out that we would have a maximum (saturation)
value for $\Gamma$.

The issue of localization in graphene has recently attracted some
attention and the chiral nature of electron behavior has been discussed
in the literature.~\cite{suzuura,mccann}
Suzuura and Ando~\cite{suzuura}
claimed that the quantum correction to the conductivity in graphene
can differ from what is observed in normal 2DEG due to the nature of
elastic scattering in graphene possibly changing the sign of
the localization correction and turn weak localization into weak
antilocalization for the region when intervalley scattering time is
much larger than the phase coherence time. Further consideration of
the behavior of the quantum correction to the conductivity in
graphene~\cite{mccann} conclude that this behavior is entirely
suppressed due to time-reversal symmetry breaking of electronic
states around each degenerate valley.

We have found through our calculations that $\Gamma$ increases
with increasing $n_i/n$ as a function of $\alpha_{gr}$. Figure~1
shows $\Gamma$ for various scattering mechanisms. As it is clear,
CDP is the dominant mechanism for $\Gamma$ in our model. The
effect of SRP is mostly negligible, except at large values of the
coupling constant. This finding is to be contrasted with
the the statement of Martin {\it et al}.\cite{martin} that both
substrate induced structural distortions (SRP) and  chemical doping
(CDP) are conceivable sources of density fluctuations. We stress
that our model calculations indicate that at realistic coupling
constant values (c.f. Fig.\,1) only the charged impurity scattering
dominates.

\begin{figure}[ht]
\begin{center}
\includegraphics[width=0.75\linewidth]{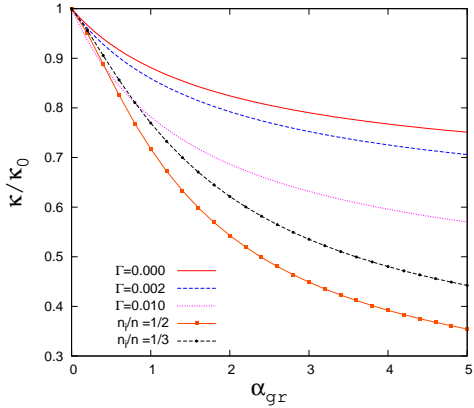}
\caption{(Color online) The compressibility $\kappa/\kappa_0$
scaled by that of a noninteracting clean system as a function of
the coupling constant $\alpha_{gr}$ for cut-off value
$\Lambda=50$.}
\end{center}
\end{figure}

\begin{figure}[ht]
\begin{center}
\includegraphics[width=0.75\linewidth]{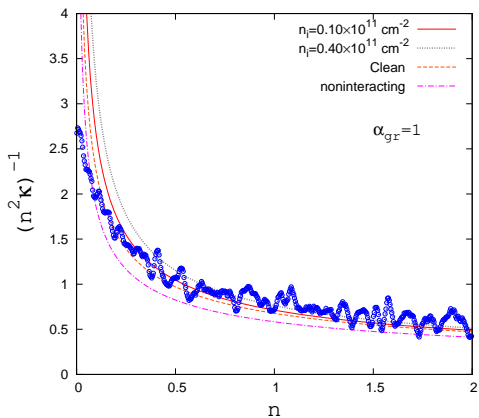}
\caption{(Color online) The inverse compressibility
$[n^2\kappa]^{-1}=\partial \mu/\partial n$ (in units of ${\rm
meV}~10^{-10}{\rm cm}^{2}$) as a function of the electron density
(in units of $10^{12}~{\rm cm}^{-2}$). The filled squares are the
experimental data by Martin {\it et al.}~\cite{martin}.}
\end{center}
\end{figure}

We have calculated the exchange and correlation energies as a
function of $\alpha_{gr}$ in the presence of disorder. It is found
that the disorder effects become more appreciable at large
coupling constants, within the mode coupling approximation. The
exchange energy is positive~\cite{yafis} because our
regularization procedure implicitly selects the chemical potential
of undoped graphene as the zero of energy; doping either occupies
quasiparticle states with positive energies or empties
quasiparticles with negative energies. Figure~2(a) shows the
correlation energy $\delta \varepsilon_c$ as a function of
$\alpha_{gr}$. It appears that the disorder effects become more
appreciable at large coupling constant. Note that $\delta
\varepsilon_c$ has the same density dependence as $\delta
\varepsilon_x$ apart from the weak dependence on $\Lambda$. In
contrast to the exchange energy, Figure~2(b), the correlation energy is
negative~\cite{yafis}. Figure~3 shows the charge compressibility,
$\kappa/\kappa_0$ scaled by its non-interacting contribution as a
function of $\alpha_{gr}$ for various models of $\Gamma$. The
behavior of $\kappa$ shows some novel physics, which is
qualitatively different from the physics known in the conventional
2DEG. Exchange makes a positive contribution to the inverse
compressibility and thus tends to reduce (rather than enhance) the
compressibility. On the other hand, correlations make a negative
contribution to the inverse compressibility and thus tends to
enhance $\kappa$. In the conventional 2DEG both contributions tend
to enhance the compressibility. In the case of graphene instead,
apparently exchange and correlation compete with each
other~\cite{martin} in determining the compressibility of the
system. It is interesting to note that similar physics is true
also in the spin susceptibility~\cite{yafis}.

In Fig.~4 we compare
our theoretical predictions for the inverse compressibility of
doped graphene with the experimental results of Martin {\it et
al.}~\cite{martin}. For definiteness we take $\Lambda=k_c/k_F$ to
be such that $\pi (\Lambda k_F)^2=(2\pi)^2/{\cal A}_0$, where
${\cal A}_0=3\sqrt{3} a^2_0/2$ is the area of the unit cell in the
honeycomb lattice, with $a_0 \simeq 1.42$~\AA~the carbon-carbon
distance. With this choice $\Lambda\simeq{(g
n^{-1}\sqrt{3}/9.09)^{1/2}} \times 10^2$, where $n$ is the
electron density in units of $10^{12}~{\rm cm}^{-2}$. Martin {\it
et al.}~\cite{martin} fitted the experimental inverse
compressibility, $(n^2 \kappa)^{-1}$ to the kinetic term using a
single parameter Fermi velocity which is larger than the bare
Fermi velocity. Note that the kinetic term in graphene has the
same density dependence as the leading exchange and correlations
terms.

As it is clear in Fig.~4 the inverse compressibility of
noninteracting system is below the experimental data. By
increasing the interaction effects, i.e., increasing the coupling
constant strength, $\alpha_{gr}$ our theoretical results move up.
Unfortunately, in the experimental sample, the value of
$\alpha_{gr}$ is not specified and we considered it to be $\approx
1$. Therefore, including the exchange-correlation effects in our
RPA theory, gives results very close to experimental data.
Furthermore, the results of incorporating impurity density, $n_i=
10^{10}$\,cm$^{-2}$ in the system and solving the self-consistent
equations to obtain the scattering rate value, yield very good
agreement with the measured values in the large and middle
electron density regions. We have examined the inverse
compressibility by the kinetic term contribution only including
a fitting value for Fermi velocity and our numerical results are well
described by a fitting velocity about $1.28\,v_F$. We would stress
here that this fitting velocity is different from
the renormalized velocity defined within the Landau-Fermi liquid
theory in graphene.~\cite{polini}

In a recent calculation of $\partial\mu/\partial n$ within the Hartre-Fock
approximation in graphene
where $\mu$ is the chemical potential and $n$ is the electron density
Hwang {\it et al}.\cite{hwang_dmu} stated that correlation and disorder
effects would only introduce small corrections.
This is not, in general, true since it has been shown by
Barlas {\it et al}.\cite{yafis}
that the correlation effects are essential in the ground-state properties.
Although these effects are not significant in very weak interaction strength
regime and high electron density, including many-body exchange-correlation
effects together with disorder effect are necessary to get agreement
with quantities measured in experiments of Martin {\it et al}.\cite{martin}
It would be useful to carry out further experimental work at larger
interaction strengths to assess the role played by correlation effects.

\section{Conclusion}

We have studied the ground state thermodynamic properties of
a graphene sheet within the random phase approximation incorporating
the impurities in the system. Our approach is based on a
self-consistent calculation between impurity effect and
many-body electron-electron interaction. We have used a model
surface roughness potential together with the charged disorder
potential in the system. Our calculations of inverse
compressibility compared with recent experimental results of
Martin {\it et al}.\cite{martin} demonstrate the importance of
including correlation effects together with disorder effects
correctly in the thermodynamic quantities.

We remark that in a very small density region, the system is highly
inhomogeneous where experimental data tends to a constant and the
effect of the impurities are very essential. A model going beyond
the RPA is necessary to account for increasing correlation effects
at low density. To describe the experimental data in this region
more sophisticated theoretical methods which incorporate
inhomogeneities are needed. One approach would be the
density-functional theory where Dirac electrons in the presence of
impurities are considered.

\begin{acknowledgments}
We thank J. Martin for providing us with their experimental data and
M. Polini for useful discussions. B.\,T. is supported by TUBITAK
(No. 106T052) and TUBA.
\end{acknowledgments}


\begin{thebibliography}{99}
\bibitem{note}
The local chemical potential is changed infinitesimally due to
impurity effect and therefore in the case of undoped graphene we
have small $\delta \mu$ around Dirac point. In case, scattering
process occurs in both conduction and valance bands near the Dirac
point and the rate of change of one-body density matrix due to collisions
is approximated by
$(d\rho/dt)_{coll}=-\gamma(\rho-\rho^c_0)-\gamma(\rho-\rho^v_0)$.
We assume that $\rho^c_0=\rho^v_0$.
\end{thebibliography}
\end{document}